\documentclass{aa}  

\usepackage{graphicx}
\usepackage{txfonts}
\usepackage{tikz}
\usepackage{sidecap}
        \sidecaptionvpos{figure}{c} 

\usepackage{subfigure}

\renewcommand{\d}[0]{{\rm d}}
\newcommand{\ind}[1]{_{\rm #1}}

\begin{document} 

\title{Cold electrons at comet 67P/Churyumov-Gerasimenko}
        
        
   \author{     I.A.D. Engelhardt               \inst{1,2}
          \and
                A.I. Eriksson                   \inst{1}
          \and
                E. Vigren                               \inst{1}
          \and
                        X. Valli\'{e}res                \inst{3}
          \and 
                M. Rubin                                \inst{4}
          \and
                        N. Gilet                                \inst{3}
          \and
                P. Henri                                \inst{3}     
                }

        \institute{
                Swedish Institute of Space Physics, Box 537, SE-751 21 Uppsala, Sweden 
         \and
             Department of Physics and Astronomy, Uppsala University, Box 516, SE-75120, Sweden
         \and
                        LPC2E, CNRS, Universit\'{e} d'Orl\'{e}ans, F-45100 Orl\'{e}ans, France
         \and
                        Physikalisches Institut, University of Bern, CH-3012 Bern, Switzerland
        }



        
\abstract
    {The electron temperature of the plasma is one important aspect of the environment. Electrons created by photoionization or impact ionization of atmospheric gas have energies $\sim$10~eV. In an active comet coma, the gas density is high enough for rapid cooling of the electron gas to the neutral gas temperature (a few hundred kelvin). How cooling evolves in less active comets has not been studied before. }
    {We aim to investigate how electron cooling varied as comet 67P/Churyumov-Gerasimenko changed its activity by three orders of magnitude during the Rosetta mission.}
     {We used in situ data from the Rosetta plasma and neutral gas sensors. By combining Langmuir probe bias voltage sweeps and mutual impedance probe measurements, we determined at which time cold electrons formed at least 25\% of the total electron density. We compared the results to what is expected from simple models of electron cooling, using the observed neutral gas density as input.} 
    {We demonstrate that the slope of the Langmuir probe sweep can be used as a proxy for the presence of cold electrons. We show statistics of cold electron observations over the two-year mission period. We find cold electrons at lower activity than expected by a simple model based on free radial expansion and continuous loss of electron energy. Cold electrons are seen mainly when the gas density indicates that an exobase may have formed.}
    {Collisional cooling of electrons following a radial outward path is not sufficient to explain the observations. We suggest that the ambipolar electric field keeps electrons in the inner coma for a much longer time, giving them time to dissipate energy by collisions with the neutrals. We conclude that better models are required to describe the plasma environment of comets. They need to include at least two populations of electrons and the ambipolar field.}


        
\maketitle

        
\section{Introduction}

The spacecraft Rosetta orbited the nucleus of comet 67P/Churyumov-Gerasimenko (hereafter 67P) beginning in August~2014 (3.6~AU heliocentric distance), passing perihelion (1.24~AU) in August~2015, and ending the mission at 3.83~AU on September~30, 2016. During all this time, the plasma environment and its evolution were observed by the instruments of the Rosetta Plasma Consortium \citep[RPC,][]{Carr2007}. As the comet approached the Sun, the nucleus was heated and neutral gas sublimated from its surface. The gas molecules can then be ionized, mainly through photoionization and electron impact ionization \citep{Vigren2016a,Galand2016a,Heritier2017b}. The electrons that are created in this process have a temperature of approximately 10~eV \citep{Haberli1996a,Vigren2013a}. 
Via frequent collisions, the electron temperature, $T_e$, can approach the neutral gas temperature, $T_n$, which according to models as well as observations is a few hundred kelvin \citep[$\gtrsim 0.01$~eV,][]{Tenishev2008,Biver2015}. This requires that the neutral gas density is sufficiently high for frequent electron-neutral collisions. These e-n collisions cool down the electrons as part of their energy is transferred into internal excitation of the molecules.

To quantify what is meant by sufficiently high neutral density, the concept of an electron exobase can be used, also known as the collisionopause or cooling boundary \citep{Mandt2016}. This is defined as the distance where the electron neutral mean free path is equal to the distance to the nucleus (see also Section~\ref{sec:exobaseModel}). Electron cooling is expected to be efficient mostly inside this distance. The exobase is not a sharp boundary, but a useful characteristic distance to indicate how much electron cooling occurs. For highly active comets such as comet 1P/Halley, which was visited by the Giotto and other spacecraft at 1~AU in 1986, the exobase was expected to be far away from the nucleus and electron cooling therefore very efficient in the inner coma. Giotto was unable to directly observe cold electrons, but a plasma density change 15~000~km from the nucleus was interpreted as indirect evidence of an electron collisionopause \citep{Ip1986a,Gan1990a,Haberli1996a}. 

\citet{Eriksson2017} reported the first direct observations of cold electrons in the inner coma of a comet by use of the Langmuir probe instrument \citep[RPC-LAP, ][]{Eriksson2006} on board Rosetta. \citet{Gilet2017} also observed cold electrons at 67P using an independent technique \citep[the mutual impedance probe MIP, ][]{Trotignon2007}.

There have been no Rosetta reports of a plasma with only cold electrons. The spacecraft potential was typically 5 -- 15~V negative during the full mission \citep{Odelstad2015,Odelstad2017} with exceptions only in low-density plasmas. This is attributed to charging of the spacecraft by the $~10$~eV electrons originating from ionization of the neutral gas before they have had time to cool. The Langmuir probe sweeps with cold electrons shown by \citet{Eriksson2017} all showed negative spacecraft potential of this order and some also direct signatures of warm electrons. In addition, the MIP spectra reported by \citet{Gilet2017} showed that warm electrons were present even when the cold electrons were seen. In a recent study of the diamagnetic cavity, \citet{Odelstad2018a} found cold electrons in at least 96\% of all LAP observations in the diamagnetic cavity, while data taken immediately outside the cavity did not always show signs of the cold population. The spacecraft potential was strongly negative in all cases, also indicating a warm electron population. \citet{Mandt2016} showed that Rosetta was almost always outside the electron exobase, which explains why warm electrons were seen all the time.

\citet{Eriksson2017} and \citet{Gilet2017} only showed a few examples of cold electron observations. 
In this paper we use a similar method to obtain statistics of cold electron observations during the complete 25~month Rosetta mission at comet 67P. To interpret the statistics, we compare results to two simple models of electron cooling. The paper is organized in the following way. In Section~\ref{sec:theory} we introduce the models. Section~\ref{sec:method} describes the instrumentation, data, and analysis methods. We present the observations in Section~\ref{sec:results} and discuss our conclusions in Section~\ref{sec:conclusions}.


\section{Theory}
\label{sec:theory}

This section describes two simple models of electron cooling and the electron cooling boundary. We use them in Sections~\ref{sec:results} and \ref{sec:conclusions} to interpret the observations.

\subsection{Sharp exobase model}
\label{sec:exobaseModel}
 

In this model, electron cooling is assumed to be efficient only inside the electron exobase, which is a region of gradual transition and can be seen as a characteristic scale length outside where the electrons are no longer collisional. Therefore it indicates at which time we should expect to see cold electrons. \citet{Eriksson2017} used such a model where all electrons created inside the exobase were assumed to be cold, so that no cooling occurred outside this boundary, and the plasma outside was a mix of cold and warm electrons depending on the distance to the exobase. Here we consider whether the activity is sufficiently high for any collisional region to form. This requires that the nominal exobase distance, calculated from observed neutral gas density, is outside the radius of the comet nucleus, which we take to be $R = 2$~km. If this is the case, we may expect a cold electron population, but not otherwise.

The electron exobase, or collisionopause or cooling boundary, is defined as the distance $r = L_c$ to the center of the nucleus where the electron mean free path $\lambda$ is equal to the neutral gas density scale height, defined as $H = n_n/(\d n_n/\d r)$. In the expanding comet atmosphere, the neutral gas density decays as $n_n\sim 1/r^2$, even when the atmosphere is not spherically symmetric \citep{Tenishev2008,Hansen2016}. This means that the scale height is $H \sim r$, so that the exobase is defined by $r = \lambda(r)$ and thus
\begin{equation}\label{eq:Lc}
        L_c = n_n(r)\sigma r^2
,\end{equation}
where r is the cometocentric distance of Rosetta, $n_n$ is the neutral density, and $\sigma$ the electron-neutral collisional cross section. Following \citet{Mandt2016}, \citet{Eriksson2017}, and \citet{Henri2017}, we use $\sigma = 5\cdot10^{-20}cm^2$ for 5~eV electrons colliding with H$_2$O molecules. 

The spacecraft position in units of $L_c$ indicates how collisional the electrons are at the current position of Rosetta, 
\begin{equation}
        R^* = \frac{r}{L_c} = \frac{1}{n_n\sigma r}.
\end{equation}
No collisional cooling is expected when the electron exobase distance $L_c$ is less than the nucleus size of about 2~km since this means that no collisional region formed.
During most of the time, Rosetta is outside this boundary, $R^*>1$ \citep[Fig.\ 5]{Mandt2016}. This indicates that there is less local collisionality, but collisions may still occur farther away from the nucleus (higher value of $R^*$).

\begin{figure}
        \includegraphics[width=\columnwidth]{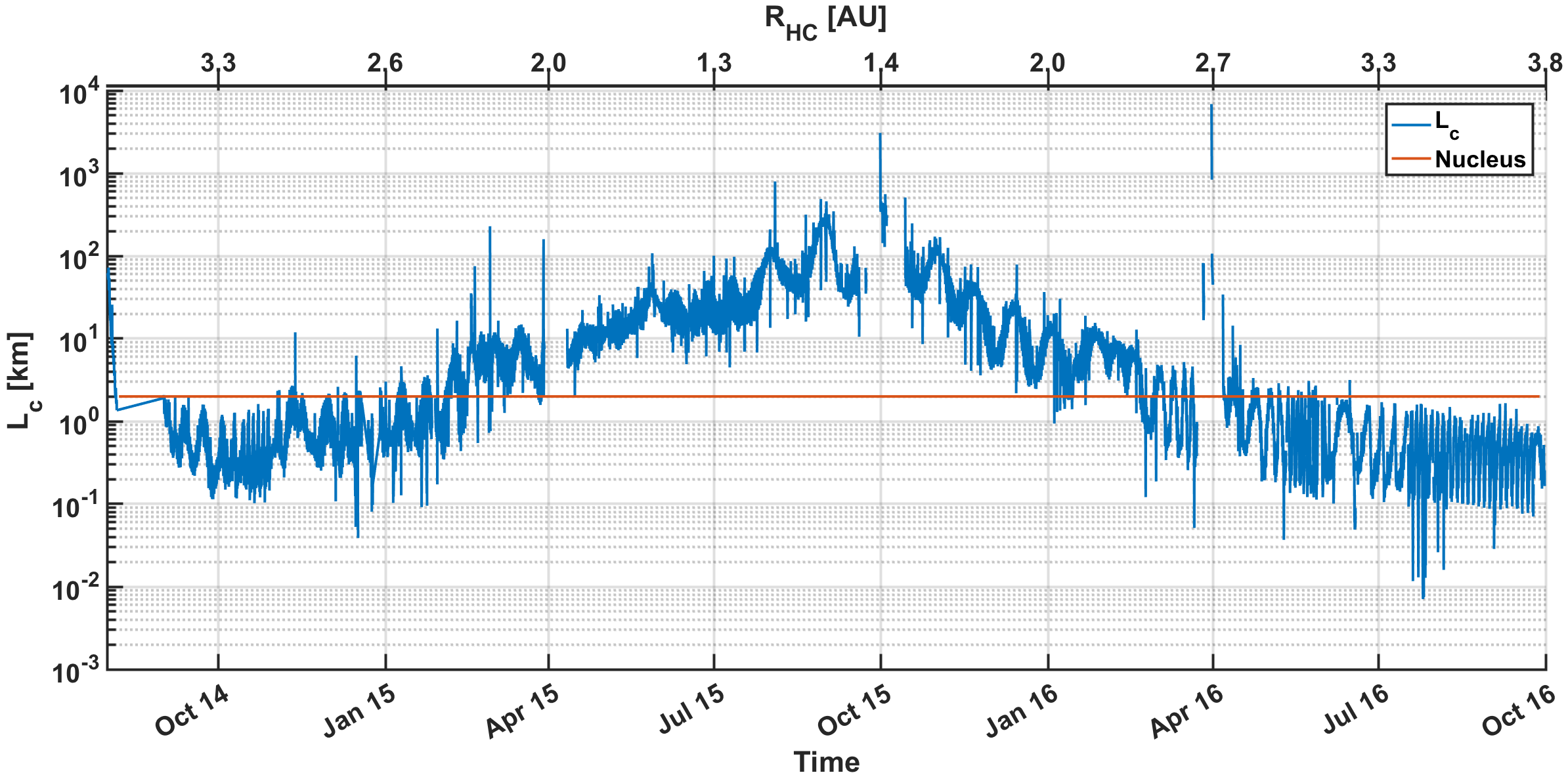}
    \caption{Calculated electron exobase distance, $L_c$ as derived from COPS (blue) for the whole mission. The red line marks the 2~km radius of the nucleus. The top legend indicates the heliocentric distance in AU. An exobase inside the nucleus is meaningless, therefore significant electron cooling is expected only when the blue line lies above the red.}
    \label{fig:Lc}
\end{figure}

Figure \ref{fig:Lc} shows the calculated electron exobase position during the whole mission, obtained from the ROSINA COmet Pressure Sensor (COPS), which measured the total density of volatiles at the location of the Rosetta spacecraft (for details, see section \ref{sec:ROSINA}).
The red line gives a nucleus radius of 2~km. Based on this, cold electrons are expected to be detectable between approximately March 2015 and March 2016. This limit is discussed further in section \ref{sec:missionOverview}.

The electron exobase depends largely on the outgassing rate, which for a spherically symmetric gas flow at constant speed is given by
\begin{equation}\label{eq:Q}
        Q = 4 \pi r^2 n_n u
\end{equation}
and can be expressed with the electron exobase distance as
\begin{equation}
        Q = 4\pi u L_c / \sigma
,\end{equation}
where $u$ is the neutral gas radial velocity.
Setting $L_c$ to 2~km for the nucleus radius, and using a radial velocity of about 1000~m/s, a minimum outgassing rate $Q > 5\cdot10^{26}s^{-1}$ is required to form an electron exobase inside which electrons can be considered collisional. For 67P, this corresponds to a heliocentric distance of about 2.3~AU \citep{Hansen2016}, therefore we do not expect cold electrons to be seen by Rosetta outside this range. It can be noted that the value of the neutral speed only affects the conversion of measured density to production rate. All our results are independent of the neutral gas velocity.

This does not mean that the model predicts that Rosetta should have seen cold electrons inside 2.3~AU because the formation of a collisional region does not hold information on how large the portion of cold electrons is at some point outside it. Assuming all electrons created inside the exobase are cooled, that no cooling occurs outside the exobase, and that the electrons follow the neutral gas flow radially outward, \citet{Eriksson2017} showed that the portion of cold electrons at cometocentric distance $r$ should be $r/L_c$. Depending on how large a portion of the electrons need to be cold for Rosetta to have observed them, this means that it may have been only much closer to the Sun than 2.3~AU that cold electrons started to be observed in Rosetta data even if this simplified model is applicable.

\subsection{Continuous cooling model}
\label{sec:coolingModel}

In this model we calculate the average energy lost by an electron due to collisions with the neutrals as function of distance. We assume that the electron moves radially outward with no change of travel direction in the collisions. 
In reality, the electron motion will be influenced by the electric- and magnetic field environment and angular scattering, which means that our model underestimates the distance traveled by the electrons.
The particle tracings by \citet{Munoz2008} show that the angular deviations for many electrons are not very large, presumably because of the strong propensity for forward scattering \citep[Fig.~5]{Itikawa2005}. The model is run without effects on the electron motion by electric and magnetic fields. This is a critical simplifying assumption that we discuss further in section \ref{sec:conclusions}. 

\begin{figure}
        \centering
    \subfigure[]{
    \includegraphics[width=\columnwidth]{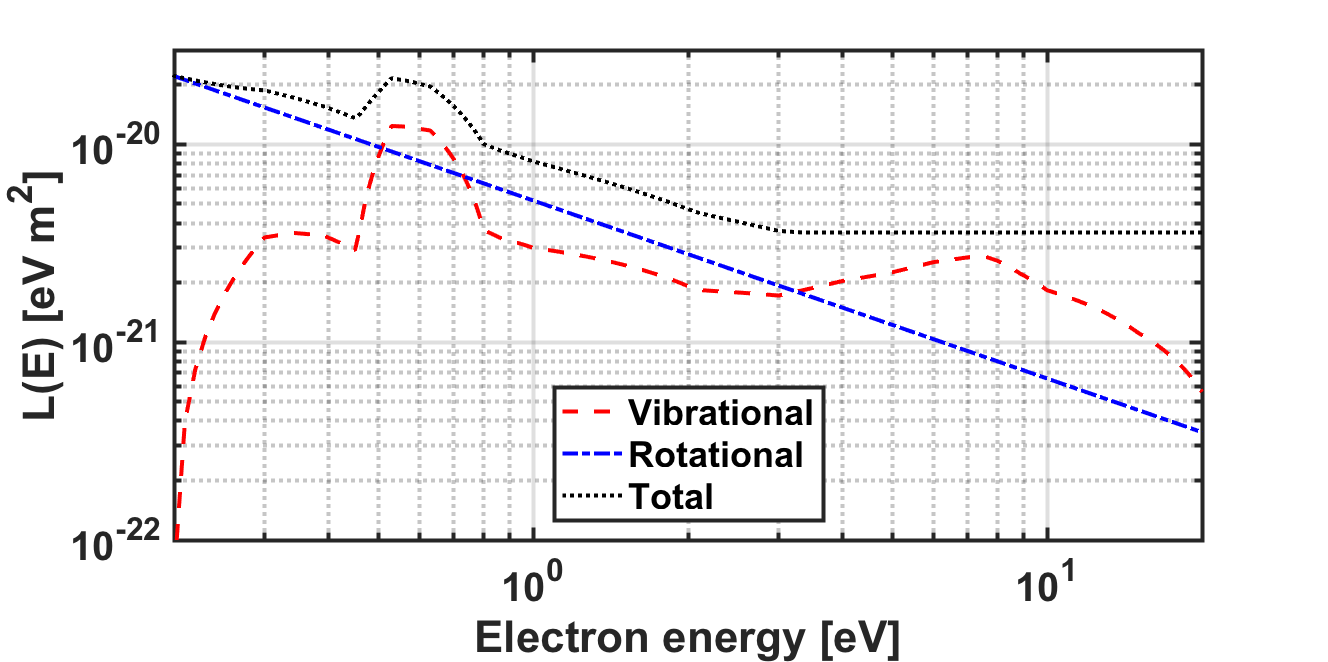}
    \label{fig:eloss_a}
    }
    \subfigure[]{
    \includegraphics[width=\columnwidth]{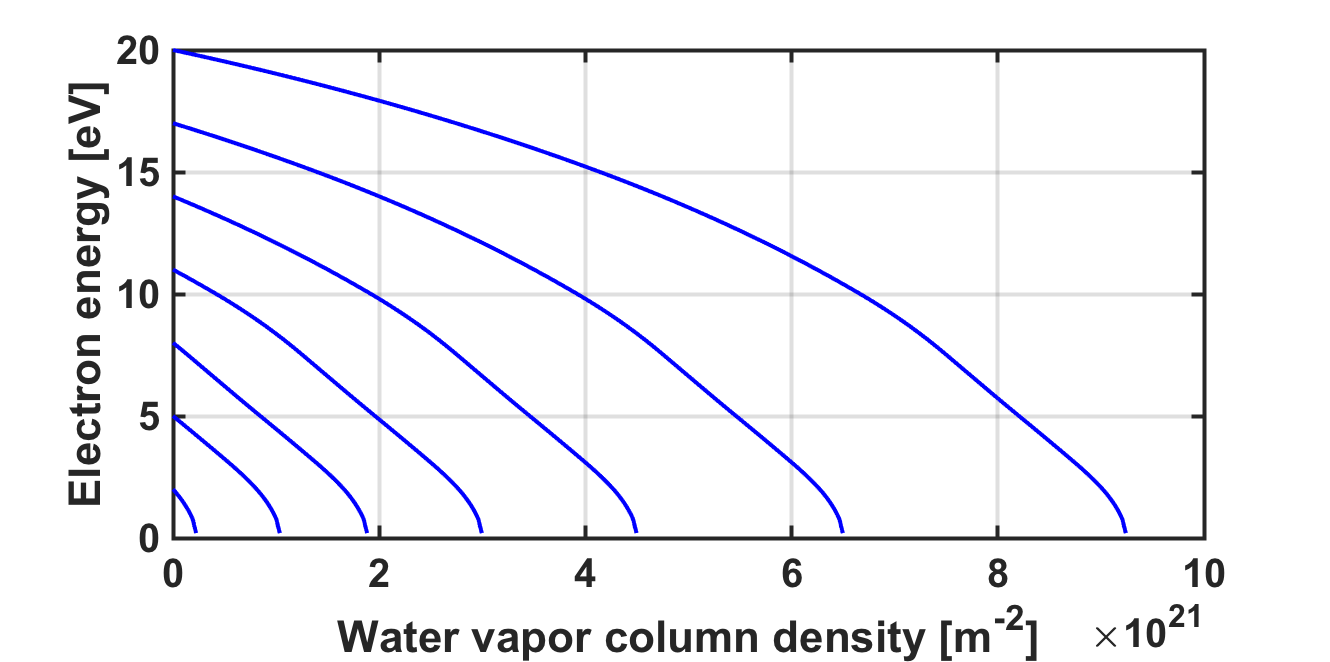}
    \label{fig:eloss_b}
    }
    \subfigure[]{
    \includegraphics[width=\columnwidth]{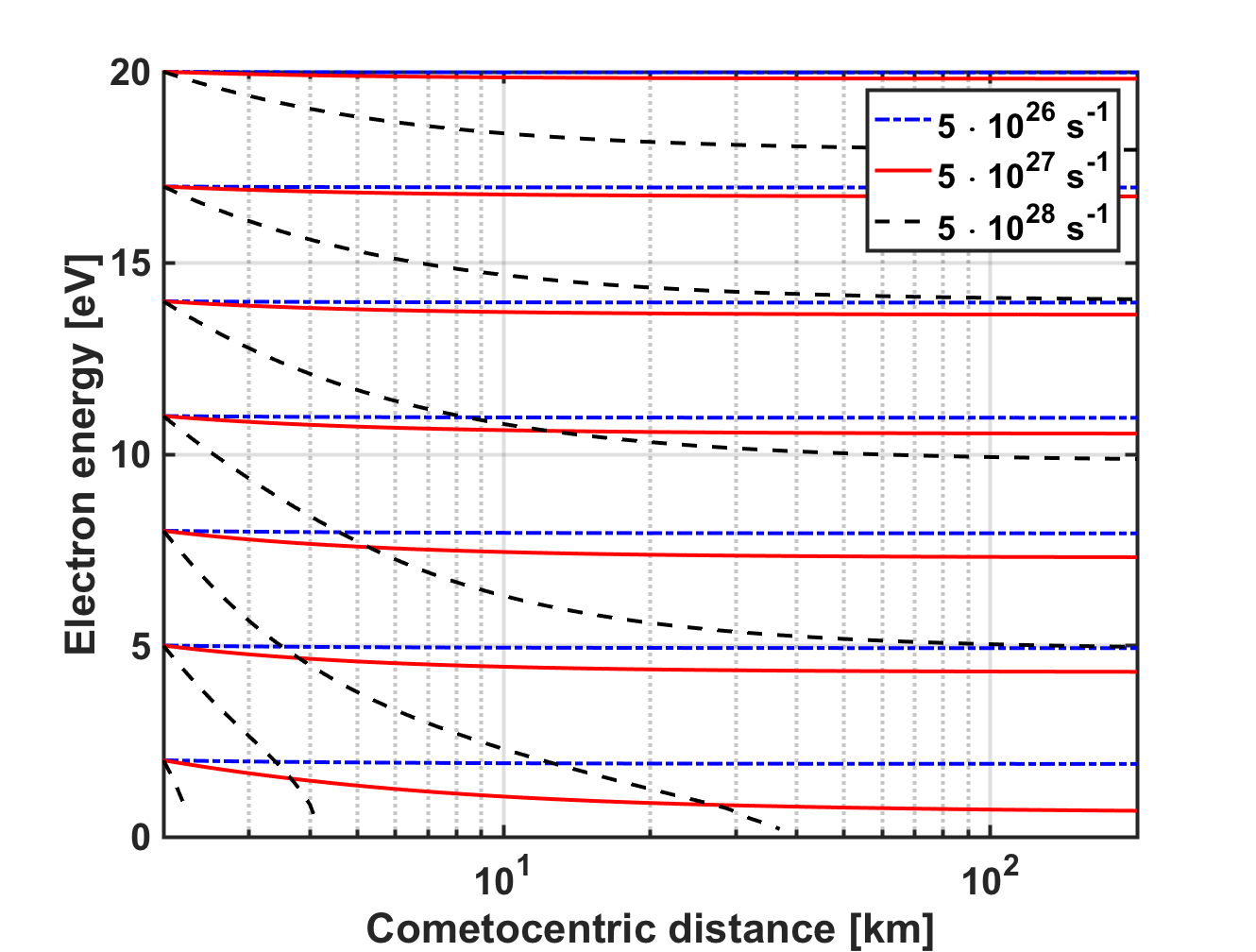}
    \label{fig:eloss_c}
    }
        \caption{Energy loss for electrons in water vapor at 300~K for electron energies between 0.2 and 20~eV. \ref{fig:eloss_a} Energy loss function. \ref{fig:eloss_b} Electron energy as function of column density from the starting point for various initial energies. \ref{fig:eloss_c} Electron energy as function of distance for electrons starting at the nucleus with various initial energies for three values of the production rate $Q$. Rotational loss is approximated by Eq.~\ref{eq:L(E)}. Vibrational loss is calculated from Eq.~(11) of \citet{Cravens1986} with cross sections from Table~9 of \citet{Itikawa2005}. The vibrational loss reaches zero at the lowest threshold energy of 0.198~eV.}
        \label{fig:L}
\end{figure}

The average energy change by an electron with kinetic energy $E$ when traveling a distance $\d l$ in a neutral gas of density $n\ind{n}$ can be written in terms of the electron energy loss function $L(E)$ as
\begin{align}\label{eq:dE}
        \d E = - L(E)\, n_n\, \d l.
\end{align}
For our needs, it is sufficient to consider the electron energy range $0.2 \leq E \leq 20$~eV. The interaction of an electron with a neutral gas molecule, which we here assume to be H$_2$O, results in a change of the translational and rotational motion of the molecule, or a change in vibration or electronic excitation state. 

\citet{Cravens1986} calculated the electron energy loss in water vapor by use of molecular calculations and experimental data. For  electrons with energy $E \gtrsim 0.2$~eV, they found that a good approximation for the loss function from excitation of rotational degrees of freedom of the water molecules is 
\begin{align}\label{eq:L(E)}
        L\ind{rot}(E) = k\ \left( \frac{1\ {\mathrm eV}}{E} \right)^{0.9},
\end{align}
where $k = 5.2 \cdot 10^{-21}$~eV~m$^2$. For the loss function that is due to vibrational excitations, we use Eq.equation~(11) from \citet{Cravens1986}, but take the required cross sections from the recommended values in Table~9 of \citet{Itikawa2005}, which covers electron energies up to 20~eV. The neutral gas temperature is set to 300~K. The resulting vibrational loss function $L\ind{vib}(E)$ is plotted in Figure~\ref{fig:L}(a) together with the rotational loss function $L\ind{rot}(E)$ from Eq.~\ref{eq:L(E)} and their sum $L(E)$. As we discuss at the end of this section, we do not include cooling by electronic excitations of the water molecules.


If we assume the electrons move radially outward, $\d l$ in Eq.~\ref{eq:dE} can be replaced by $\d r$, so that
\begin{equation}\label{eq:diff}
        \frac{\d E}{L(E)} = -n_n(r)\, \d r
,\end{equation}
which can be integrated to relate the energy $E$ at position $r$ of an electron starting at energy $E_0$ to the column density $N(r_0,r)$ between the start and end points:
\begin{equation}\label{eq:int}
        \int_{E}^{E_0} \frac{\d E'}{L(E')} = \int_{r_0}^r n_{n}(r')\ \d r' = N(r_0,r).
\end{equation}
Panel~(b) of Figure~\ref{fig:L} shows the result of numerically integrating the energy integral in Eq.~\ref{eq:int} with the $L(E)$ plotted in panel~(a). Electrons starting at various energies (at the left of the plot) can be seen to lose energy continuously with increasing column density as they move through the gas. As expected from the decay of the loss function with energy in panel~(a), electrons at low energy lose energy more quickly.

Assuming radial motion in the $1/r^2$ gas density profile given by Eq.~\ref{eq:Q}, the column density (Eq.~\ref{eq:int}) can be expressed in terms of $Q$ or the locally observed neutral gas density $n\ind{n}(r)$ as 
\begin{equation}\label{eq:col}
        N(r_0,r) = \frac{Q}{4 \pi\, u} \left[ \frac{1}{r_0}-\frac{1}{r} \right] = r\,   n_{n}(r)\, \left[ \frac{r}{r_0}-1 \right].
\end{equation}

As we assume radial motion, the electrons having lost most energy should be those created close to the nucleus. Setting $r_0$ to the nucleus radius $R = 2$~km, $E(r)$ as calculated from Eqs.~\ref{eq:int} and \ref{eq:col} therefore becomes the lowest energy we expect any electron with initial energy $E_0$ to have when reaching $r$. Figure~\ref{fig:L}(c) plots energy as function of distance out to 200~km for electrons starting at the nucleus ($r = R =2$~km) for three different production rates: $Q = 5 \cdot 10^{26}$~s$^{-1}$ (red), $Q = 5 \cdot 10^{27}$~s$^{-1}$ (blue), and $Q = 5 \cdot 10^{28}$~s$^{-1}$ (black). 

In the low-activity case $Q = 5 \cdot 10^{26}$~s$^{-1}$ , none of the tracked electrons suffer a significant energy loss. 
This activity level was typical of 67P six months before and after perihelion, at about 2.3~AU \citep{Hansen2016}. It is also the production rate that in Section~\ref{sec:exobaseModel} was considered to be the minimum for the formation of a collisional region inside an electron exobase. The formation of this region at this low activity is not supported by the continuous cooling model. 

The medium activity production rate $Q = 5 \cdot 10^{27}$~s$^{-1}$ is typical for 67P about two months before and three months after perihelion (about 1.4 and 1.6~AU, respectively). It shows appreciable cooling of at least the lowest energy electrons (red solid line in Figure~\ref{fig:L}c). 

The high-activity production rate $Q = 5 \cdot 10^{28}$~s$^{-1}$ was reached by 67P only close to perihelion. The black dashed lines in Figure~\ref{fig:L}c show significant energy loss for all the tracked electrons; an electron with an initial high energy of 8~eV reached 0.2~eV (where our model ends) already at a cometocentric distance of 40~km.

\citet{Vigren2013a} show that above $\sim 7$~eV, electron energy loss by electronic excitation of the water molecules may be important. \citet{Itikawa2005} noted that lack of good experimental data for this process complicates detailed modeling. To determine how important our neglect of this can be, we also performed all calculations with a modified model, where we replaced the value of $L(E)$ for $E > 7$~eV with $L(7$~eV$)$. While some adjustment could be seen in the plots for more than 10~eV initial energy in Figure~\ref{fig:L}(b) and (c), changes in the comparisons to observational data we present in Section~\ref{sec:missionOverview} were very marginal and do not affect any conclusions. In the remaining paper, electronic excitation is therefore neglected, although we note that improvement on this point could be desirable for detailed event studies, for example.
Based on this model, a cold electron population is not expected to be observed by Rosetta at heliocentric distances $\gtrsim 1.5$~AU.


\section{Instrumentation and method}
\label{sec:method}

\subsection{RPC-LAP}
\label{sec:RPC-LAP}
The main data were acquired from the Langmuir probe instrument \citep[LAP, ][]{Eriksson2006} included in the Rosetta Plasma Consortium \citep[RPC, ][]{Carr2007}. LAP includes two identical Langmuir probes, LAP1 and LAP2, which are situated on booms. We used only LAP1 here. It is pointed toward the typical direction of the nucleus and therefore sees fewer spacecraft perturbations in the plasma than LAP2.
The operational mode used was a sweep through voltages where the probe measures the resulting current. These sweeps were usually at 160~s intervals, sometimes even 64~s. They rarely occurred at other intervals, but they were always a multiple of 32~s. The data were binned into non-overlapping two-hour intervals. This means a a maximum of 112.5 sweeps every two hours (64~s), but the majority was 45 sweeps every two hours (160~s).

\subsection{RPC-MIP}
\label{sec:RPC-MIP}
The MIP instrument \citep{Trotignon2007} provided the plasma density we used to derive the temperature. It observed the response to an emitted signal at different frequency steps. The electron plasma density was inferred from on-ground identification of the resonance at the plasma frequency, which directly gives the plasma density \citep{Gilet2017}. The plasma density has to be above or around 100~cm$^{-3}$ (the limit slightly depends on the electron temperature) to provide a reliable measurement.

\subsection{ROSINA-COPS}
\label{sec:ROSINA}
From the Rosetta Orbiter Spectrometer for Ion and Neutral Analysis \citep[ROSINA, ][]{Balsiger2007}, we used the total neutral gas density from the nude gauge of the comet pressure sensor (COPS). We used the COPS data for model calculations of the expected electron cooling.

The density measured by COPS was not only due to the gases that are naturally present in the coma, but it was also influenced by various gas sources on the spacecraft \citep{TzouThesis}. Most important of these are dust impact signatures, daily but brief thruster firings for reaction wheel off-loading, rarer but longer thruster firings for orbit changes, and perturbations associated with changing solar illumination on the spacecraft surfaces as the pointing changes \citep{TzouThesis}. For some pointing, the spacecraft blocked COPS from access to the coma gas flow \citep{Odelstad2018a}, resulting in artificially low neutral densities in the data set. COPS was turned off during orbit correction maneuvers. Most of the remaining spurious effects were shorter than 10~minutes. We have used the median value over 30~minutes of COPS data. This removes most contamination events from the COPS signal. Some may be left, but not so many as to significantly skew the statistics.

\subsection{Electron temperature estimate}
\label{sec:elTemp}

For a positive probe, the Langmuir probe current $I$ is mainly due to electrons attracted from the plasma. This current is dependent on the plasma density, $n_e$, and the electron temperature, $T_e$, 
\begin{equation}
        I = 4 \pi\, a^2 n_e \sqrt{\frac{e T_e}{2 \pi\, m_e}\,} \left( 1 + \frac{U}{T_e}  \right)
,\end{equation}
where $a = 0.025$~m is the radius of the probe, $e = 1.6\cdot10^{-19}$~C is the electron charge, $T_e$ is in eV, $U$ is the probe potential with respect to the plasma, and $m_e=9.1\cdot10^{-31}$~kg is the electron mass. In the actual LAP sweeps, the bias voltage $V_b$ was set and varied, but $V_b$ differs from $U$ only by the spacecraft potential $V_s$, which because of the hugely different surface areas of the spacecraft and probe can be taken to be constant during a sweep. The slope $S={dI}/{dU}$ can therefore be directly determined from the recorded sweeps with little error, and 
\begin{equation}\label{eq:slope}
        S = \frac{dI}{dU} = a^2e^{3/2}\sqrt{\frac{8\pi}{m_e}}\frac{n_e}{\sqrt{T_e}}.
\end{equation}
If we have a value for $n_e$, we can therefore derive an estimate of the cold electron temperature from $S$ through Eq.~\ref{eq:slope}. Analysis of the LAP sweeps can provide values for $n_e$ as well as $T_e$. This is sometimes possible also when there are two populations \citep{Eriksson2017}, although no reliable automatic procedure for this case exists and the uncertainties may be large. We instead used the independent density measurement by MIP. The main cost of this is that plasmas with densities below 100~cm$^{-3}$ cannot be investigated. As we expect cold electrons mainly when the density is high, we do not consider this to be a problem, and it will in any case not cause any false detections of cold electrons. 

Combining observed LAP $S$ and MIP $n_e$, we find a temperature estimate from Eq. \ref{eq:slope} as
\begin{equation}\label{eq:Te}
        T_s = 8 \pi\, \frac{a^4 e^3}{m_e} \, \frac{n_e^2}{S^2}.
\end{equation}
It should be noted that the value $T_s$ we obtain from this method equals the mean kinetic energy of the particles only for a Maxwellian distribution. For a sum of two Maxwellians, the calculated temperature will be closer to the temperature of the colder of the two distributions. The reason is that the slope $S$ is mainly determined by the cold population, as discussed above. For example, a mix of 50\% cold electrons at 0.03~eV (300~K) and 50\% warm photoelectrons at 10~eV would result in a $T_s$ value of 0.1~eV from Eq.~\ref{eq:Te}.

The plasma density can vary greatly between sweeps \citep{Eriksson2017,Engelhardt2018a}, therefore to calculate $T_s$ , we only used coincident data from LAP and MIP. The two instruments were synchronized so that when both were operating, an MIP spectrum was taken within a few seconds of an LAP sweep. However, since MIP spectra did not always result in a density value, mostly because the density was too low or too high for the operational mode used, $T_s$ cannot be derived for all sweeps. Of the 386~581 LAP1 sweeps available during the mission, 119~172 have simultaneous MIP measurements within 3~seconds of the end time of the LAP sweep. The end time of the sweep is relevant as we used the last data points in each sweep to derive the slope $S$ (see next paragraph). 
These are then the only ones we used for this study. Thus every sweep we analyzed has simultaneous MIP measurements.
 
We derived electron slopes from sweeps on LAP1 in the following way. First we cut away any points with currents above 9.5~$\mu$A to ensure that we avoided any effects of the instrument electronics saturation, which lay at about 9.8~$\mu$A. Then we took the remaining five points with the highest bias voltage (or as many points needed to span at least 1~V) and determined the slope by a linear least-squares fit to these points. This was done to ensure that we selected the steepest part of the probe curve, see for example Figs. 2 and 6 in \citet{Eriksson2017}. The sweeps extended from negative to positive bias voltage, therefore we used the last data points in every sweep.

Equation~\ref{eq:Te} is quadratic both in the LAP slope and the MIP density. This means that the $T_s$ estimate is sensitive to errors in both of these quantities. The error in $n_e$ from MIP is considered to mainly be the random errors caused by finite frequency resolution and incorrect identification of the plasma resonance in the spectrum, this does not add any systematic bias to the statistics. For LAP, the random numerical error in determining the slope by the method described above should be small. However, it is possible that the electric field from the negatively charged spacecraft repels electrons so that some or all of the lowest energy electrons cannot reach the probe \citep{Olson2010}. This then gives an underestimated slope and a too high $T_s$ estimate. The contribution to the slope from the warm population has the same effect: it causes us to overestimate the temperature of the cold population.


\section{Results}
\label{sec:results}

\subsection{Cold electron identification}
\label{sec:slope}

\begin{figure}
        \centering
    \includegraphics[width=\columnwidth]{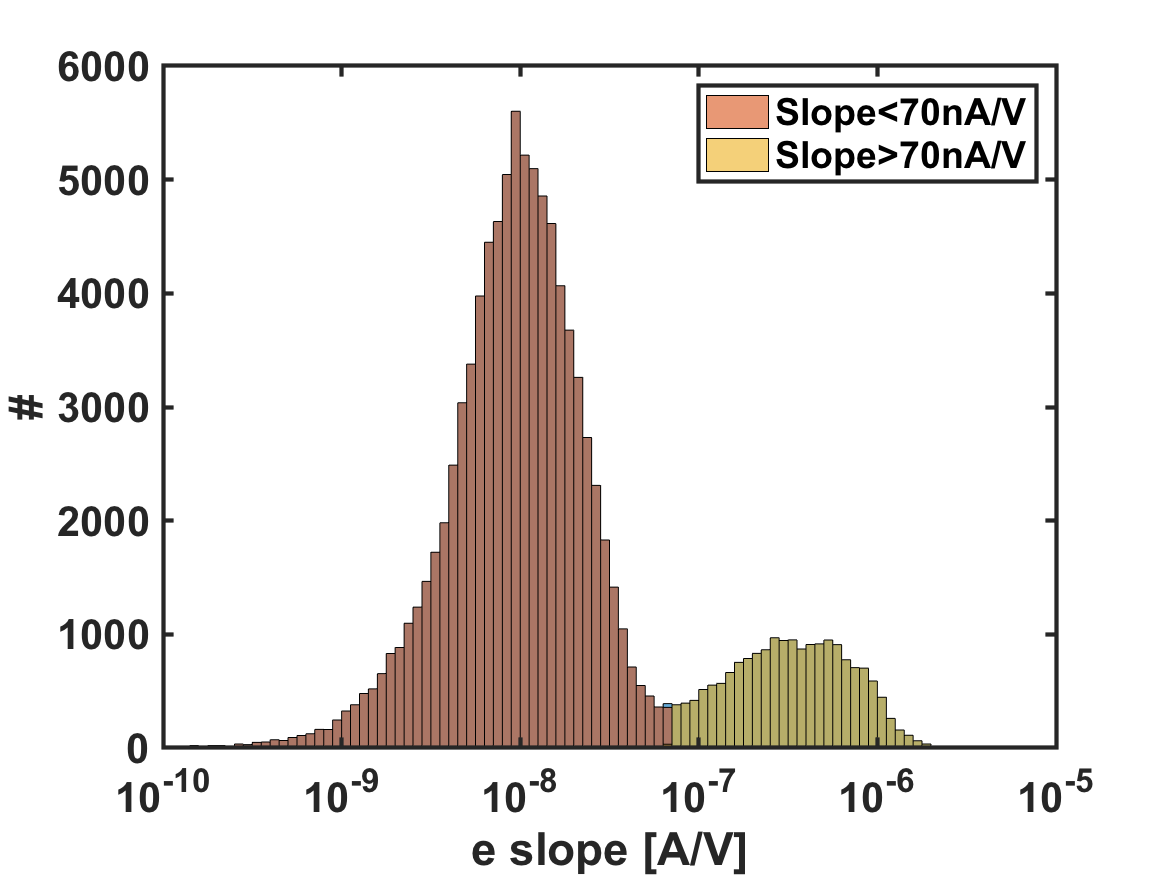}
        \caption{Histogram of the slopes of all suitable Langmuir probe data from August 7, 2014, to September 30, 2016. The yellow bars show the data classified as steep slopes (>70nA/V), while the red bars represent the shallow slopes.}
    \label{fig:SlopeStats}
\end{figure}

\begin{figure}
                \includegraphics[width=\columnwidth]{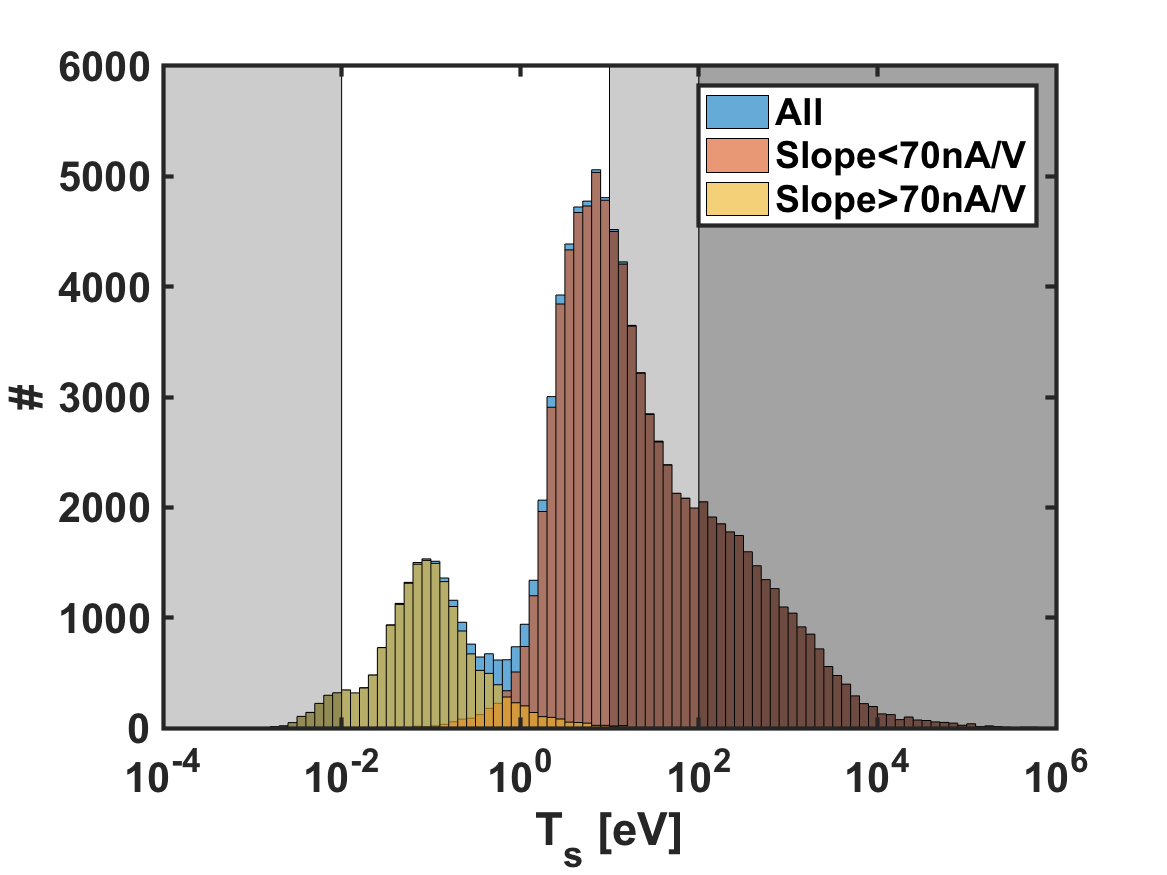}
        \caption{
    Histogram of electron temperature derived from LAP1 and MIP from the entire Rosetta mission at the comet, August 7,~2014, through September 30,~2016. The yellow and red bars indicate values derived from sweeps with a steep and shallow slope as defined in Fig.~\ref{fig:SlopeStats}. Individual data points may have large errors, and the parts of the distributions below 0.01~eV and above 10~eV are not reliable.  
    }
        \label{fig:histoTe}
\end{figure}

Figure~\ref{fig:SlopeStats} shows a histogram of the electron slope from all LAP1 sweeps between the arrival of Rosetta within 100~km from 67P on August 7, 2014, and the end of the mission on September 30, 2016. The data peak around 10~nA/V. However, a second peak at about $500$~nA/V exists, and the minimum around 70~nA/V separates the data set into two distinct populations. We define slopes higher than 70~nA/V as steep (color-coded in yellow), and slopes lower than this value as shallow (red).

Figure~\ref{fig:histoTe} shows histograms of the electron temperature calculated through Eq.~\ref{eq:Te} from the 119~172 simultaneous LAP1 slopes and MIP densities. As before, the red data come from the slopes that are less than 70~nA/V, and the yellow data come from the steep slopes. Like in Figure~\ref{fig:SlopeStats}, the data are separated into two populations. One group of sweeps peaks around 0.1~eV and the other around 10~eV, with a minimum in the histogram around 1~eV. The color code shows that the low $T_s$ group corresponds to the steep slopes and the higher $T_s$ group to the shallow slopes. The derived temperature values suggest that we interpret the two groups as plasmas with and without cold electrons, respectively. We recall that in the presence of two populations, the calculated temperature is mainly determined by the temperature of the cold population.

As discussed in Section~\ref{sec:elTemp}, the $T_s$ estimate is sensitive to errors in $n_e$ and $S,$ which means that Figure~\ref{fig:histoTe} must be interpreted with caution. In particular, no $T_s$ value above the main peak at 10~eV should be trusted. High electron fluxes with energies of a few hundred eV were often observed on Rosetta \citep{Clark2015a,Broiles2016b}, but since the method we used is more sensitive to low energies, $T_s$ values of hundreds of eV or more would require the bulk plasma to have this temperature. This would have charged the spacecraft to hundreds of volts, which is not observed in the data. Data from LAP and the ion composition analyzer (RPC-ICA) show that the occurrence of such charging levels are  very rare \citep{Odelstad2017,StenbergWieser2017a} and the large number of sweeps with $T_s\gg 10$~eV is therefore an artifact. Any effect of the negative spacecraft potential blocking electron access to LAP1 would give a slope that would be too low and hence too high $T_s$ \citep{Olson2010}.

Nevertheless, the key properties of the histogram are reasonable. The main peak around 10~eV corresponds to what is expected for photoelectrons \citep{Vigren2013a} as well as to LAP bias sweep fits \citep{Eriksson2017}. The second peak around 0.1~eV could be expected for cold electrons. These should not be colder than the surrounding neutral gas, which means that the left shoulder of this distribution below 0.01~eV is probably spurious. However, most values are very reasonable as an effective temperature of a mix of cold and warm electrons.

\begin{figure*}
        \centering
        \includegraphics[height = 0.5\textheight]{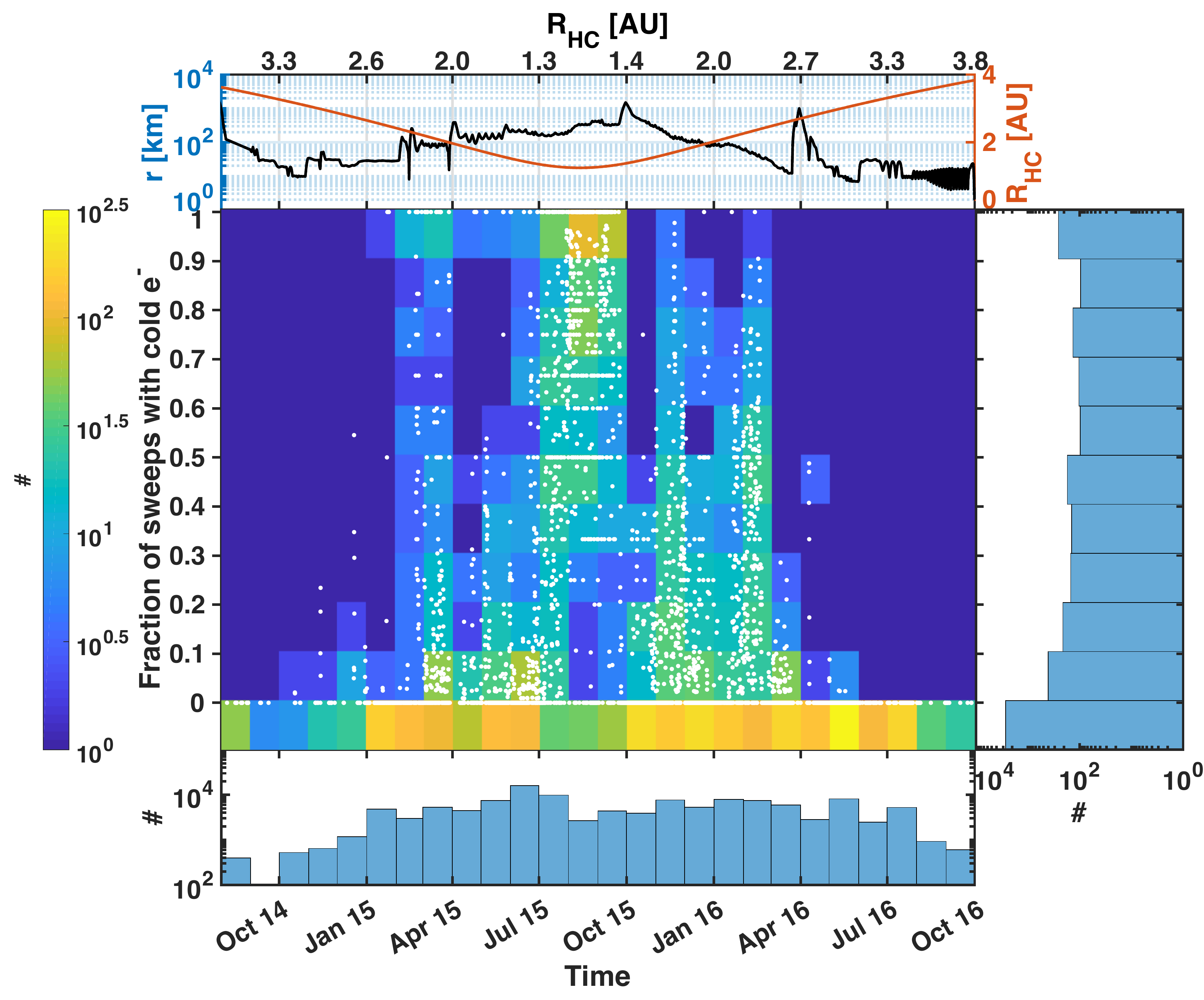}
    \caption{
    Mission overview of the fraction of sweeps containing cold electrons ($T_s<0.3$~eV and $S >70$~nA/V). The top panel shows the radial distance of Rosetta to the nucleus and the heliocentric distance of the comet. The white data points in the main panel show the fraction of sweeps containing cold electrons for every two-hour interval. The colored background shows a 2D histogram representation of the data. The horizontal bins are per month. Tick marks show the first of the indicated month. The vertical bins are the fraction of sweeps during a two-hour interval that contains cold electrons, each bin spanning 10\%. For every month, the color code then gives the total number of two-hour intervals with cold electrons identified in the given fraction of sweeps. The total number of two-hour intervals per month is not constant, as there must be useful LAP and MIP data for at least one data point per two-hour interval. The histograms below and at right show the total bin count.}
    \label{fig:Te_steep_Full}
\end{figure*}

To summarize, Figures~\ref{fig:SlopeStats} and \ref{fig:histoTe} indicate that the LAP sweeps separate into two categories depending on whether the slope on the electron side is steep or shallow. Including MIP data in the analysis shows that these groups correspond to $T_s$ values around 0.1 and 10~eV, respectively. As $T_s$ mostly depends on the colder electron population, we assumed that the sweeps with steep slopes to show a cold electron population. To minimize the amount of false cold electron detections, the criteria we used for a cold electron population for the remaining study is that (i) the LAP electron slope is steep, (ii) there is an MIP density measurement within 2~s of the LAP sweep, and (iii) the combined LAP-MIP temperature estimate through Eq.~\ref{eq:Te} is lower than 0.3~eV.

We did not attempt to measure the fractions of the electron density that is due to warm and cold electrons. However, we can set a lower limit to the relative abundance of cold electrons. If there are two electron populations at temperatures $T_c$ and $T_w$ with relative contributions to density of $\alpha$ and $1-\alpha$, the temperature $T_s$ we derive from Eq.~\ref{eq:Te} will, because of Eq.~\ref{eq:slope}, be given by
\begin{equation}
        \frac{1}{\sqrt{T_s}} = \frac{\alpha}{\sqrt{T_c}} + \frac{1-\alpha}{\sqrt{T_w}}.
\end{equation}
With reasonable values of $T_c = 0.03$~eV and $T_w = 10$~eV, the cold electrons must make up at least a fraction $\alpha = 28$\% of the density in order to give $T_s < 0.3$~eV. A conservative statement is that when cold electrons are detected by the criteria above, they contribute at least on the order of 25\% to the total electron number density. If the presence of the negatively charged spacecraft influences the measurements by blocking low energy electrons from reaching the probe \citep{Olson2010}, then this should have more effect on the cold than on the warm population. It may therefore well be that cold electron fractions as low as 25\% cannot be detected. For this reason, we cannot rule out that there may be some amount of cold electrons present at the times when only shallow sweeps are observed. For this study, we therefore only claim that cold electrons make up at least 25\% of the electron density in at least the events we detect by the method above.

\subsection{Mission overview}
\label{sec:missionOverview}

\begin{figure*}
        \centering
\includegraphics[height = 0.5\textheight]{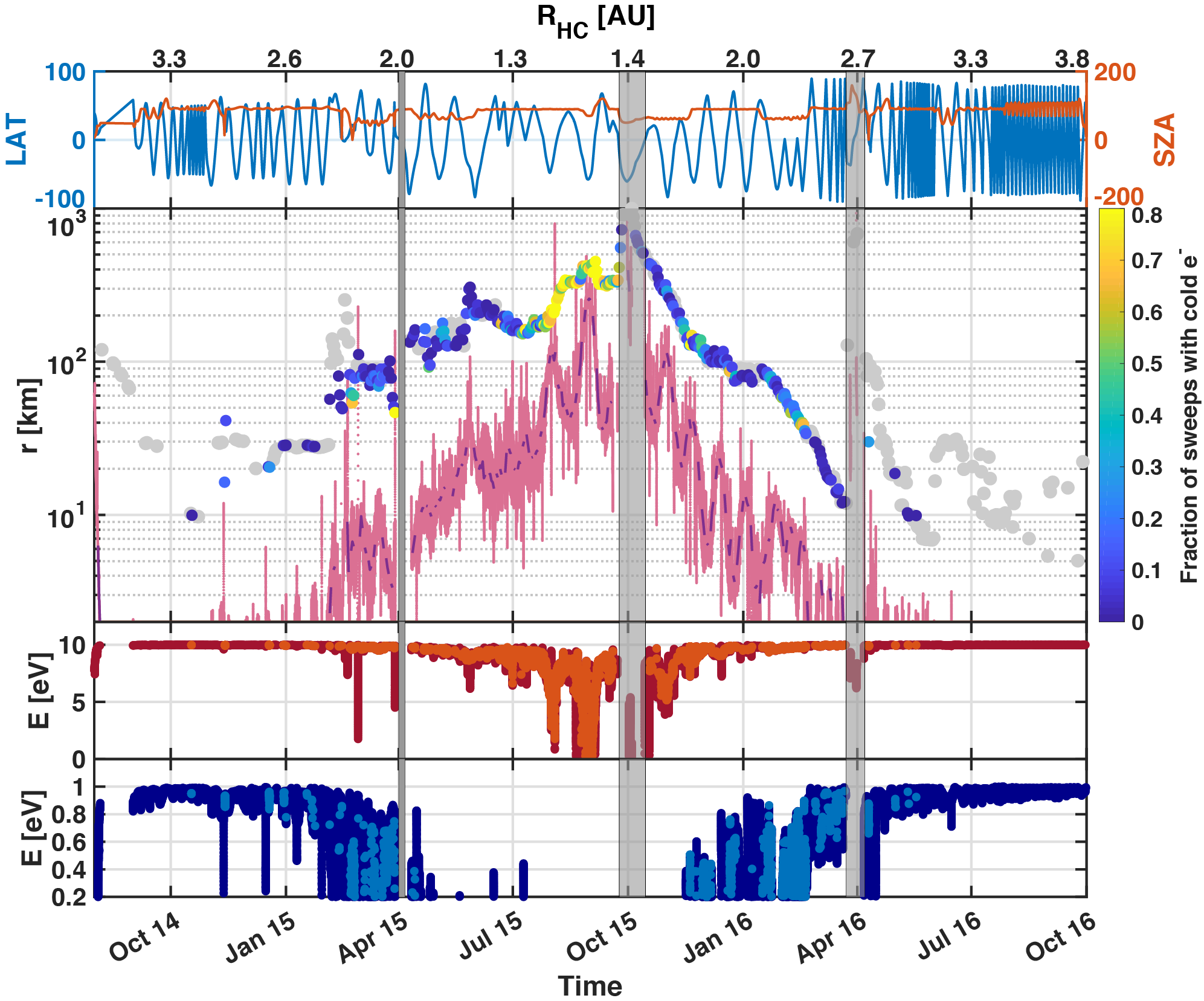}
    \caption{First panel: Latitude (blue, left) and solar zenith angle (red, right) of Rosetta. Second panel: Electron exobase distance, $L_c$ (rose), and the radial distance color-coded by the fraction of cold electrons in a 24-hour interval. When this is zero, we set the color to gray. The two lower panels show the energy a 10~eV and 1~eV electron should have at the position of Rosetta, starting at the nucleus surface. The gray overlay are excursion times.}
     \label{fig:LcRE}
\end{figure*}


Figure \ref{fig:Te_steep_Full} shows a mission overview when we detect cold electrons by the criteria set in Section \ref{sec:slope}. The top panel shows the radial distance (black line) from the center of the nucleus in logarithmic scale on the left axis. The heliocentric distance (red line) of the comet is shown on the right axis.

In the main panel the white data points show the fraction of steep sweeps showing signs of cold (<0.3eV) electrons. 
Each data point represents a two-hour interval in which we have at least one 
sweep with simultaneous MIP density so that we can search for cold electrons.
The background color plot shows the amount of two-hour intervals in the bin. The horizontal binning is one bin per month, while the vertical binning gives the fraction of sweeps with cold electrons. This is also shown in the histogram to the right of the figure. The histogram in the bottom shows the amount of data points during each month.
Figure \ref{fig:Te_steep_Full} shows that close to perihelion on August 13, 2015, LAP observed that a higher fraction of the sweeps contained cold electrons, sometimes all within a two-hour interval. 

In Figure \ref{fig:Lc} we showed the calculated exobase distance during the whole mission. In Figure \ref{fig:LcRE} we compare this to cold electron observations. The top panel shows the latitude and solar zenith angle (the angle Sun-nucleus-Rosetta, also known as the phase angle). The latitude is defined on the nucleus in the standard "Cheops" frame \citep{Preusker2015a}. In the middle panel we show the exobase distance, from the comet radius of 2~km up to 1000~km (rose points). 
Overplotted is the radial distance of Rosetta color-coded by the fraction of sweeps with cold electrons during a whole day. If we measure zero sweeps that are cold, the color is set to gray. The exobase distance describes the cold electron observations well, at least in a qualitative sense. Cold electrons are detected in the data about in the same period as the exobase is well defined ($>2$~km, November~2014 to May 2016). Furthermore, we find the highest daily fraction of sweeps with cold electrons when Rosetta was close to the electron exobase. The sharp exobase model of Section~\ref{sec:exobaseModel} therefore seems quite useful for predicting when cold electrons are present. 

The bottom two panels show the theoretical energy that an electron released close to the nucleus had at the distance of Rosetta according to the continuous cooling model presented in Section \ref{sec:coolingModel}. Here we use Eqs. \ref{eq:int} and \ref{eq:col} together with the energy loss function plotted in Figure~\ref{fig:L}(a). We set the start position $r_0$ to the nucleus radius $R =2$~km and the initial energy $E_0$ of new electrons to be 1~eV (blue) and 10~eV (red).
The darker color shows the expected energy at any time, the lighter color shows its value at the times we see cold electrons. 
There is a large variation between individual data points, which is due to variations in actual neutral gas density. The main interest is in the trend. 
Electrons starting out at 10~eV close to the nucleus clearly only lose a significant part of their energy in some events around perihelion, as expected from the discussion in Section \ref{sec:coolingModel}. 
For electrons starting out at 1~eV, the cooling is more efficient (see Figure~\ref{fig:L}a), and they could lose most of their energy in the period of February 2015 to March 2016. However, as most fresh photoelectrons have around 10~eV energy, this model suggests that cold electrons should be abundant only in the months around perihelion.

The shaded regions represent the times when the spacecraft was at unusually large cometocentric distance where the neutral gas density is low and errors in COPS data therefore grow. This means the calculated exobase distances and electron cooling should not be trusted here, but the cold electron detection is still good. These excursion periods are from April~1 to 6, 2015, from September~24 to October~15, 2015 (dayside excursion), and from March~23 to April~7, 2016 (nightside excursion).

\subsection{Location}
\label{sec:location}

In the previous section we investigated how the cold electrons evolved in time over the Rosetta mission. We here study how the cold electrons were distributed in space around the nucleus.

\begin{figure}
        \includegraphics[width=\columnwidth]{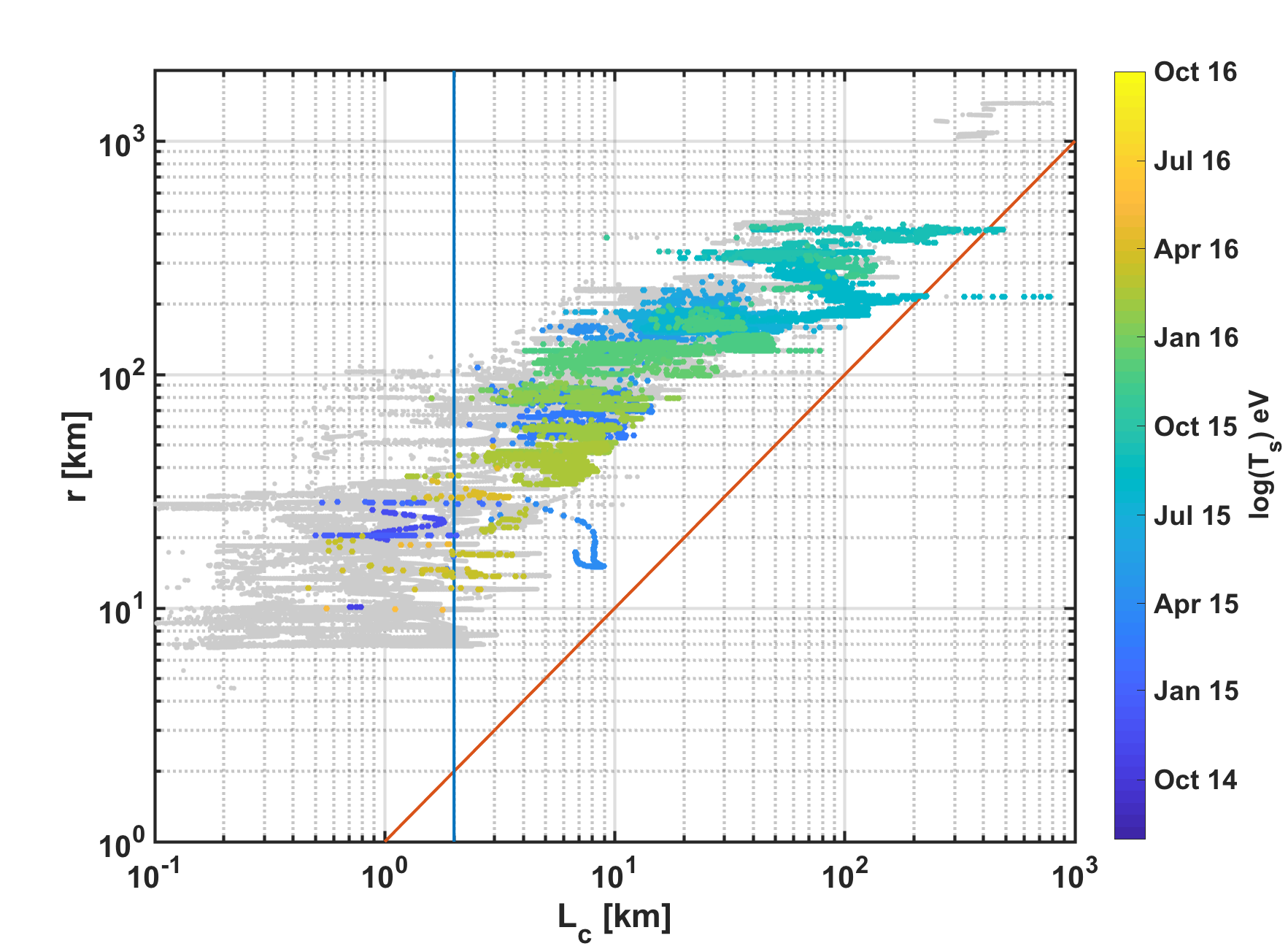}
    \caption{Radial distance vs. electron exobase distance. Colored data show the cold electrons from the steep slope (<0,3eV and >70~nA/V). Gray shows all the sweeps. The red line indicates where $L_c = r,$ and the blue line represents the 2~km distance of the nucleus radius.}
    \label{fig:ExorL}
\end{figure}

Figure \ref{fig:ExorL} shows the radial distance of Rosetta versus the calculated exobase distance. The gray dots show the data of all sweeps. The colored data shows the data from the steep slopes Te < 0.3 eV. The red line is the position where $L_c=r$. The blue line marks the minimum exobase distance of 2~km. For data points to the left of this line, there should be no region where electron collisions are important.

\begin{figure}
        \centering
    \subfigure[]{
    \includegraphics[width=\columnwidth]{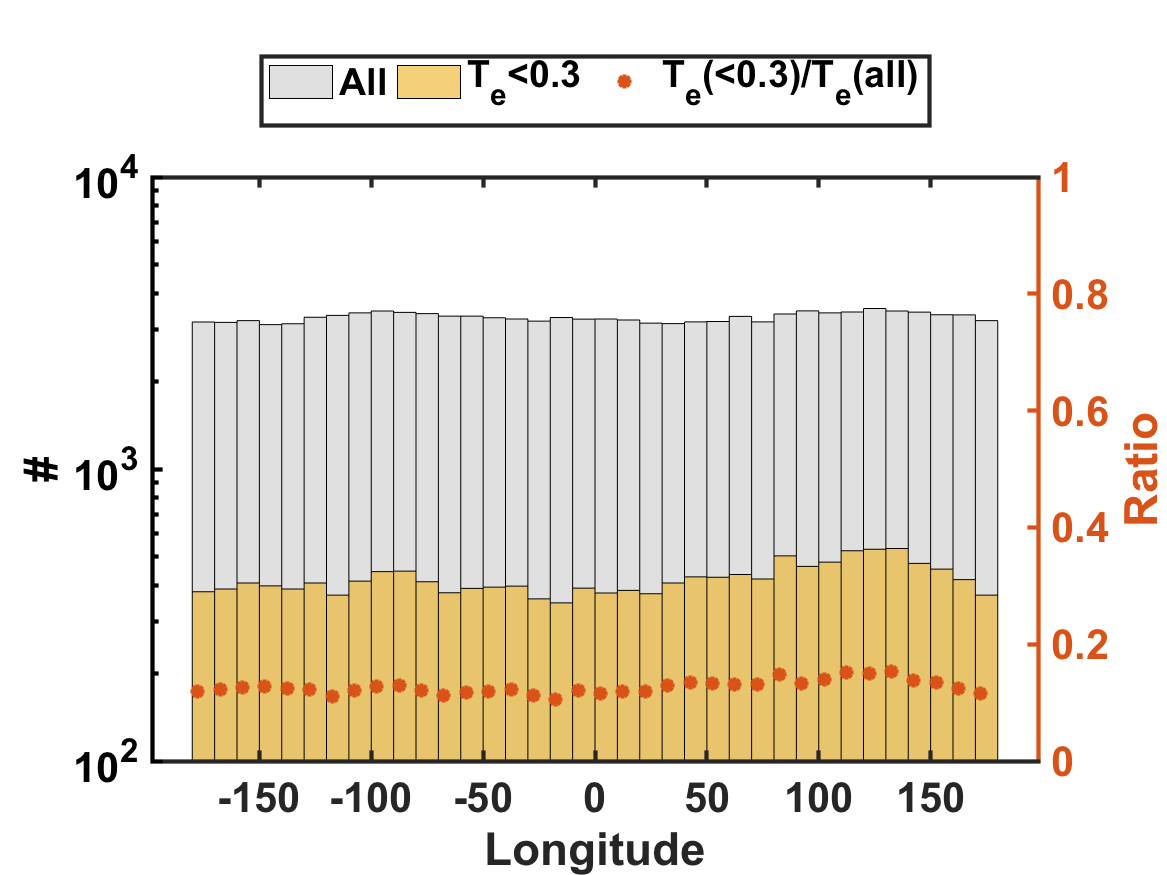}
    \label{fig:lonTe}
    }
    \subfigure[]{
    \includegraphics[width=\columnwidth]{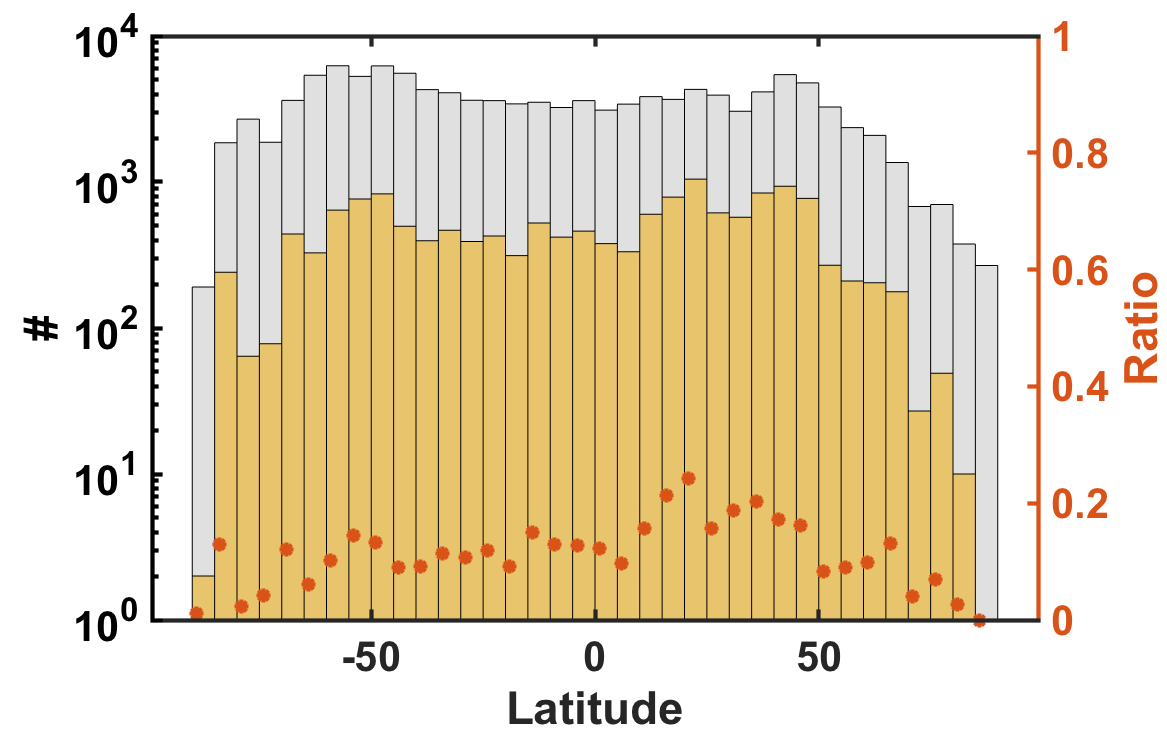}
    \label{fig:latTe}
    }    
    \caption{Histograms of all temperatures from all slopes during the whole mission (gray) and of cold electrons of steep slopes (yellow), with the ratio given (red points), and the right y-axis. 
    Panel \ref{fig:lonTe} shows this vs. longitude, and panel \ref{fig:latTe} shows it vs. latitude.
    }
    \label{fig:lonlatTe}
\end{figure}

Rosetta obviously collected most of the data outside of the exobase, with only very few exceptions. The cold electron observations (colored dots) gather to the right side of the gray cloud marking all observations, meaning closer to the red line and hence to the exobase. Interestingly, a few data points can be found to the left of the blue line, in a situation when no collisional region should exist in the sharp exobase model. A possible mechanism could be the ambipolar electric field, which we discuss in Section~\ref{sec:conclusions}. Nevertheless, there are clearly fewer events with cold electrons at low activity (low $L_c$ value), as we discussed in the previous section.

\begin{figure}
        \centering
    \includegraphics[width=\columnwidth]{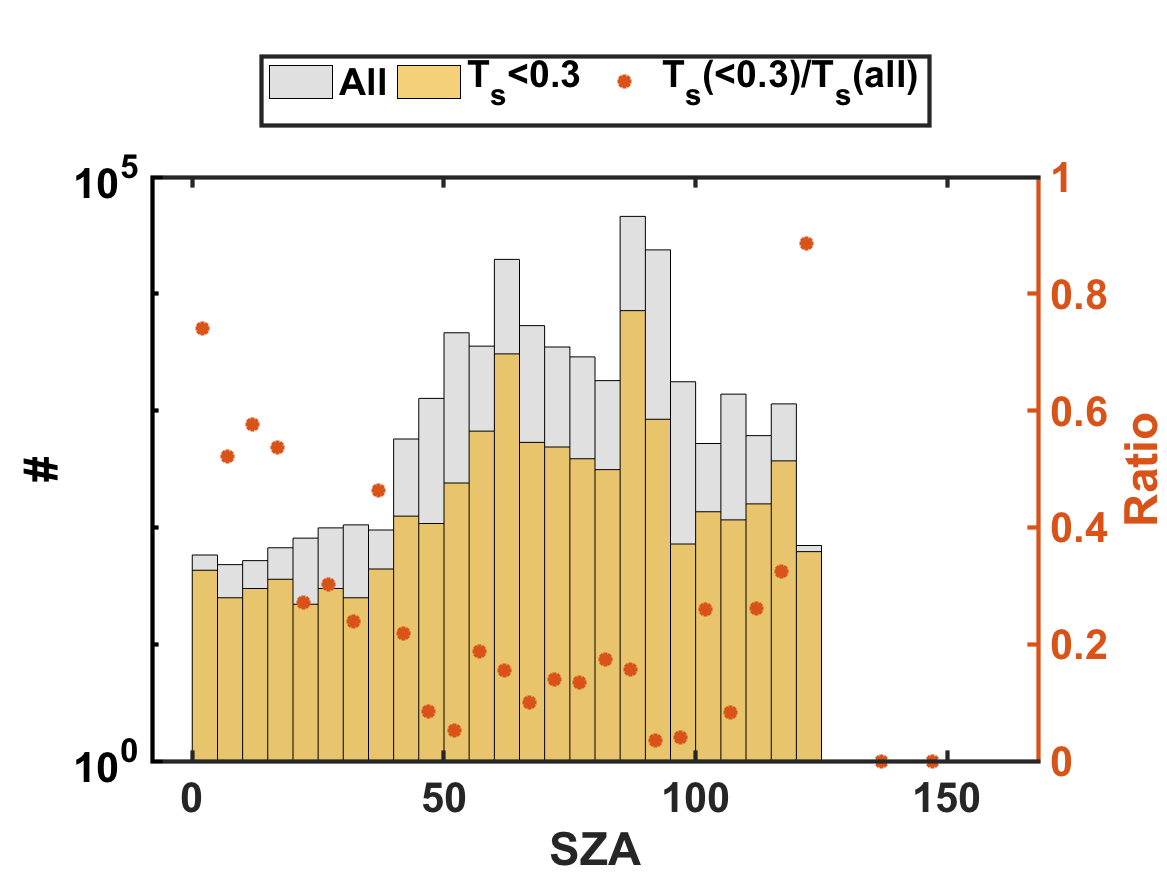}
    \caption{Gray histogram: Coverage of sweeps over the SZA over the course of the mission. Yellow histogram: Amount of cold electrons from steep slopes. The red points show the fraction of gray and yellow bins, and the right y-axis. More cold electrons are observed at low angle.}
    \label{fig:szaTe}
\end{figure}

Next we investigated if there is a relation between cold electrons and position angle around the nucleus in Figures~\ref{fig:lonlatTe} and \ref{fig:szaTe}. The histograms in these figures have the same layout: the gray bars show the number of total sweeps in each bin, while the yellow bars present the number of data points showing cold electrons, $T_e<0.3$~eV. The red points show the fraction of cold electrons versus all.

In Figure \ref{fig:lonlatTe} we show the longitude in panel \ref{fig:lonTe} and the latitude in panel \ref{fig:latTe}. Because the comet rotated beneath Rosetta, the comet longitudes are covered quite evenly, as the gray histogram of Figure \ref{fig:lonTe} shows. One small peak lies at about 120$\degr$ in the detection of cold electrons, and possibly also a small hint of a peak around $-100\degr$. This can be compared to the known distribution of neutral gas and plasma in the northern hemisphere during the period of northern summer, which lasted until the equinox in May 2015. During this period, the neck region (longitude around $\pm 100\degr$) was the most active \citep{Hassig2015a,Edberg2015a,Odelstad2015}, therefore this is also where we should expect most cooling in this period. Figures~\ref{fig:Te_steep_Full} and \ref{fig:LcRE} show, however, that most cold electron observations are made in the period after this equinox (southern summer) and then the outgassing variation with longitude was much weaker \citep{Hansen2016}. This can explain why the peaks we see over neck region longitudes in the full data set are weak.

A study of the latitude shows fewer cold electrons over the poles. This is also consistent with the known outgassing pattern, as most gas is emitted where the illumination of the nucleus is strongest. The irregular shape of the nucleus and the orbit eccentricity of 67P complicates calculation of the solar flux at any single point, but it is on average lower close to the poles than around the equator \citep{Hansen2016}.

Figure \ref{fig:szaTe} shows the cold electrons versus solar zenith angle, SZA, on the nucleus' surface. Cold electrons on the dayside (SZA$<90\degr$) are obviously more likely, peaking at $0\degr$, that is,\ on the Sun-nucleus line. This is where the neutral density is highest \citep{Hansen2016}, so that this is where the cooling should be most efficient.

The nightside excursion in March~2016 reached close to $160\degr$ , but then the distance was large so that the neutral gas density was low. Most of the time, Rosetta remained close to terminator orbit (SZA$=90\degr$), not only when the activity was high, but also in the early and late parts of the mission, for example, when the nucleus was mapped in preparation for the Philae landing in November 2014. This is probably why the percentage of sweeps showing cold electrons have a minimum around the terminator. In regular operations, SZA was always kept below $120\degr$. It is unclear why the fraction of sweeps with cold electrons have a maximum at this largest phase angle. This might be some influence of an unusual spacecraft illumination, possibly causing spacecraft outgassing that disturbed the probe measurement, and we consider this single data point as unreliable. In summary, the cold electron distributions in longitude, latitude, and solar zenith angle all are as expected based on where the neutral density is known to be high, leading to a more efficient electron cooling.


\section{Conclusions and discussion}
\label{sec:conclusions}

The first conclusion from the data presented in Section~\ref{sec:slope} is that the Langmuir probe sweep measurements can be used as a proxy for cold electrons. Using the slope $S$ of the part of the LAP sweeps where electrons are attracted together with the independent plasma density estimate by MIP, we can give an estimate of the electron temperature $T_s$. Figure~\ref{fig:histoTe} showed two peaks in the temperature distribution we derived: one at 10~eV, and one at about 0.1~eV. These are consistent with theory where newly created electrons have a temperature of about 10~eV and the neutral gas has a temperature $\gtrsim 0.01$~eV, and also with previous Rosetta observations \citep{Eriksson2017,Gilet2017}. We showed that if there are these two populations, we obtain$T_s < 0.3$~eV if the cold electrons contribute $\gtrsim25$\% of the density.
We therefore defined criteria for the presence of cold electrons to be when $T_s < 0.3$~eV and LAP sweeps have steep slopes ($S>70$~nA/V).

Our major result is that cold electrons were seen throughout most of the Rosetta mission. Figure~\ref{fig:Te_steep_Full} shows that cold electrons were regularly observed as early as mid-November~2014 and through April~2016. This means they were seen for heliocentric distances $\lesssim 3$~AU, which corresponds to a production rate $Q \gtrsim 10^{26}$~s$^{-1}$ \citep{Hansen2016}. They were commonly seen (about half the time or more) from July~2015 until Rosetta left the innermost coma for the dayside excursion in late September~2015 (Figure~\ref{fig:LcRE}), which means $Q \gtrsim 10^{28}$~s$^{-1}$.

We showed for the distribution of cold electrons in space around the nucleus that they are more likely to be seen when Rosetta was close to (or the few times inside) a nominal electron collision boundary calculated from the neutral gas density observed by COPS. 

The correlation with the solar zenith angle is strong, as we see more cold electrons at low values. 
They are also more common at low cometary latitude. All this is as expected from where the neutral gas density is highest \citep{Hansen2016}, so that electron cooling should be most efficient there. 

We find that the sharp exobase model as described in Sect. \ref{sec:exobaseModel} indicates quite well when cold electrons may be expected.
Cold electrons are expected between some time in November~2014 to March~2015 and March 2016 
(Figure \ref{fig:Lc}), which agrees with the data.

Based on the results from the continuous cooling model in Sect. \ref{sec:coolingModel}, we also expect that electrons created close to the nucleus with an energy of 1~eV should have lost most of their energy when they reached Rosetta in about this time frame. 
Most electrons are created at about 10~eV, however. For these electrons, the neutral gas density is high enough to cool them only during one to two months around perihelion.
The cooling boundary as defined by Eq. \ref{eq:Lc} therefore seems to predict better when to expect cold electrons than the more elaborate cooling model (Sect. \ref{sec:coolingModel}).

The continuous cooling model assumes that the electrons move radially outward on a straight line. This would not be a poor assumption if the only force acting on the electrons were collisions with the neutrals because the typical change in direction after the interaction with a water molecule is small \citep[the scattering cross section peaks sharply at small angles,][]{Itikawa2005}.

Therefore some other force is needed to keep the electrons near the nucleus for a sufficiently long time to cool. One mechanism is the ambipolar electric field, which is required for quasineutrality \citep{Vigren2017}.
The electrons move faster than the ions, so that in a density gradient, an electric field forms in the direction opposite to the gradient. This ambipolar electric field retains some electrons, which gives them more time to collide with the neutral gas molecules. Another cooling effect of the ambipolar field is that an electron moving against the field converts kinetic energy into potential energy, and it then leaves the vicinity of the nucleus with a lower energy. 

Collisionless particle-in-cell simulations of comet 67P at low activity by \citet{Deca2017a} showed the formation of this ambipolar electric field and that it does trap cometary electrons. However, models or simulations that also include electron cooling are not yet available. 

A continued data analysis is likewise required. We have only considered the overall statistics of cold electrons at large scales in space and time. It should be possible to learn more from detailed studies of the cold electron distribution during selected events and shorter time periods. The diamagnetic cavity has been studied recently \citep{Odelstad2018a}, but the region outside the cavity needs to be better understood as well. In a magnetic field, the cold electrons are more strongly magnetized than the warm electrons, and the magnetic field is therefore expected to be important for organizing them.

    
\begin{acknowledgements}
        The research in this paper was funded by the Swedish National Space Board under contracts 171/12, 109/12 and 166/14. Work at LPC2E/CNRS was supported by ESEP, CNES and by ANR under the financial agreement ANR-15-CE31-0009-01. Discussions within ISSI International Team on the Plasma Environment of Comet 67P after Rosetta (number~402) have been very useful. This work has made use of the AMDA and RPC Quick look database, provided by a collaboration between the Centre de Donn\'{e}s de la Physique des Plasmas (CDPP) (supported by CNRS, CNES, Observatoire de Paris and Universit\'{e} Paul Sabatier, Toulouse), and Imperial College London (supported by the UK Science and Technology Facilities Council). Rosetta is a European Space Agency (ESA) mission with contributions from its member states and the National Aeronautics and Space Administration (NASA). Work on the ROSINA COPS at the University of Bern was funded by the State of Bern, the Swiss National Science Foundation, and by the European Space Agency PRODEX program.
\end{acknowledgements}
        
        
        \bibliographystyle{aa}
        \bibliography{Manbib} 

\end{document}